\documentstyle[12pt]{article}
\begin{document}

\begin{flushright}
December 1999

OU-HET-339
\end{flushright}

\begin{center}

\vspace{5cm}
{\Large Intersecting Branes and Generalized Vortices}

\vspace{2cm}
Takao Suyama \footnote{e-mail address : suyama@funpth.phys.sci.osaka-u.ac.jp}

\vspace{1cm}

{\it Department of Physics, Graduate School of Science, Osaka University, }

{\it Toyonaka, Osaka, 560-0043, Japan}

\vspace{4cm}

{\bf Abstract} 

\end{center}

We construct the effective theory of intersecting branes and investigate the BPS monopoles 
in the theory.
The monopoles obtained are the generalization of Nielsen-Olesen vortex.
We study the properties of the solutions and interpret them as the D0-branes on the 
brane-intersections.

\newpage

{\large {\bf 1. Introduction}}

\medskip

The study of black hole physics is one of the most successful applications of string theory.
Entropy of the black hole, known as Bekenstein-Hawking entropy \cite{BH}, was derived 
from an extremal one \cite{SV}\cite{CM}, and Hawking radiation was also discussed \cite{CM}.
Similar argument was applied to 4-dimensional extremal black holes \cite{4dim1}\cite{4dim2}.
In these analyses, D-branes played the crucial role.

There are black hole solutions in supergravity which is obtained from the D-brane solutions 
and its entropy is zero.
One such example is considered in \cite{suyama}.
It is obtained from the solution which corresponds to intersecting D4-branes and 
D0-branes.
From the stringy argument, the entropy of this black hole should not be zero \cite{4dim1}.
It was pointed out that microstates which contribute to the entropy correspond to the 
monopole ground states \cite{suyama}.
Such monopoles can exist due to the appearance of the massless fields coming from the string 
stretched between the intersecting D4-branes.

In this paper we generalize the effective theory on the intersection discussed in 
\cite{suyama}\cite{intersection} to the one which describes multi-intersections, 
and investigate the properties of the BPS monopole solutions in the theory.
It will be identified with the D0-branes on the intersection.

This paper is organized as follows.
In section 2, we discuss briefly the supergravity solution and derive the entropy formula 
up to numerical factor.
In section 3, we generaize the effective theory on the intersection, and investigate the 
BPS monopole solutions.
Simple cases are discussed in detail.
Section 4 is devoted to the discussions.

\vspace{1cm}

{\large {\bf 2. Black hole entropy and intersecting branes}}

\bigskip

2.1 \ Supergravity solution

\medskip

We will consider in this paper the following brane configuration in Type IIA theory 
compactified on $T^6$.
There are $Q_1$ D4-branes wrapped along, say, (4567) directions of $T^6$, and $Q_2$ D4-branes 
wrapped along (6789) directions (we will denote this as D4'-branes).
D4-branes and D4'-branes intersect over a 3-dimensional hyperplane.
There are also $N$ D0-branes on the intersection.
This brane configuration preserves $\frac18$ supersymmetries.

The corresponding supergravity solution is known \cite{solution}.
After dimensional reduction to 4 dimentions \cite{reduction}, this is regarded as a 
black hole solution.Its metric in the Einstein frame and the dilaton field are 
respectively as follows,
\begin{eqnarray}
& & \hspace{5mm} ds_E^2  = -(1+k)^{-\frac12}(H_1H_2)^{-\frac12} dt^2+(1+k)^\frac12
                  (H_1H_2)^\frac12(dx_1^2+dx_2^2+dx_3^2) \\
& & e^{-2\phi_{(4)}} = (H_1H_2)^{-\frac16}(1+k)^\frac12 , \label{sugra}
\end{eqnarray}
where
\begin{eqnarray}
&& H_i = 1+\frac{c_iQ_i}r \; (i=1,2), \hspace{3mm} k = \frac{c_PN}r \nonumber \\
&& \hspace{1.5cm} r^2 = x_1^2+x_2^2+x_3^2 \nonumber .
\end{eqnarray}
One can see that this black hole has zero-area horizon, indicating that its 
Bekenstein-Hawking entropy is zero.

However, from the point of view of string theory, the entropy should not be zero.
The brane configuration described above is U-dual to the one which is $S^1$ compactification 
of the well-established D1-D5 system \cite{SV}\cite{CM}.
The entropy of the black hole corresponding to the D1-D5 system was derived and this 
agrees completely to the Bekenstein-Hawking entropy obtained from the supergravity solution.
The method used in \cite{SV}\cite{CM} depends only on the dyamics on the branes.
Thus the same entropy formula should be valid when one more direction is compactified.
From U-duality invariance of entropy, it is natural to expect that the entropy of the 
D4-D4'-D0 system is
\begin{equation}
S = 2\pi\sqrt{Q_1Q_2N}. \label{entropy}
\end{equation}
The emergence of the non-zero entropy is due to the following reason.
One can see that the dilaton field (\ref{sugra}) diverges at the horizon $(r=0)$.
This implies that the quantities related to the horizon will receive large quantum 
correction, and thus non-zero entropy could appear \cite{4dim1}.

\vspace{7mm}

2.2 \ D-brane description

\medskip

The low energy effective theory on the D4-D4' intersection is 3-dimensional 
${\cal N}=4$ supersymmetric gauge thoery.
The gauge group is $U(Q_1)\times U(Q_2)$.
There are three hypermultiplets: one in the adjoint representation of $U(Q_1)$, one in 
the adjoint representation of $U(Q_2)$, and one in the bi-fundamental representation of 
$U(Q_1)\times U(Q_2)$.
The D0-branes will be described as BPS monopoles, in the same way as the D0-branes are 
described as instantons which preserve half of the supersymmetries in the D4-D0 system 
\cite{BinB}.

In the simplest case, $Q_1=Q_2=1$, the adjoint hypermultiplets will decouple from the 
dynamics and the resulting effective theory is ${\cal N}=4$ supersymmetric QED with a 
charged hypermultiplet.
The bosonic part of the Lagrangian is
\begin{eqnarray}
S_{boson} &=& \int d^3x [\frac1{g^2}(-\frac14 F_{ij}F^{ij}
             -\frac12 \partial_i\phi_m\partial^i\phi_m)
             -D_iq^\dagger D^iq-D_i\tilde{q}^\dagger D^i\tilde{q} \nonumber \\
          & &-\frac{g^2}2((q^\dagger q-\tilde{q}^\dagger \tilde{q}-\zeta)^2
             +4|q\tilde{q}|^2)
             -\phi_m\phi_m(q^\dagger q+\tilde{q}^\dagger \tilde{q})] \label{effective} 
\end{eqnarray}
\begin{eqnarray}
          & &D_iq=\partial_iq-iA_iq,\hspace{3mm} 
             D_i\tilde{q}=\partial_i\tilde{q}+iA_i\tilde{q} 
               \nonumber \\
          & &(i,j=0,1,2, \hspace{2mm} m=1,2,3) , \nonumber 
\end{eqnarray}
where $\phi_m$ are the real scalars in the vector multiplet and $q,\tilde{q}$ are the 
complex scalars in the hypermultiplet. 

We have introduced the FI-parameter $\zeta$.
Since the ground states of the D4-D4'-D0 system are marginal bound states, it may be 
difficult to investigate them by the semiclassical argument.
Therefore we deformed the theory to make those states be truly bound states.
We expect that this deformation does not change drastically the properties of the monopoles.
Similar arguments have been done in \cite{intersection}.

The effective theory (\ref{effective}) has static monopole solutions known as 
Nielsen-Olesen vortex \cite{vortex}.
This solution breaks 4 supersymmetries, thus one-monopole solution has 4$(=2^2)$ ground 
states.

The entropy of the D4-D4'-D0 system can be obtaind by counting the ground states of 
$N$ monopoles.
Assume that D4(D4')-branes are separated from each other.
Then the number of ground states is equal to the number of distributing the $N$ D0-branes 
among the $2Q_1Q_2$ bosonic states and $2Q_1Q_2$ fermionic states.
This can be easily calculated and the resulting entropy is
\begin{eqnarray}
S_{stat} &=& 2\pi\sqrt{\frac16 (2Q_1Q_2+\frac12 \cdot 2Q_1Q_2)N}  \nonumber \\
         &=& \frac1{\sqrt2}\cdot 2\pi\sqrt{Q_1Q_2N}
\end{eqnarray}
This agrees with (\ref{entropy}) up to numerical factor.

\vspace{1cm}

{\large {\bf 3. Generalized vortex}}

\bigskip

Now we consider $Q_1,Q_2\geq 1$ case.
Again, for simplicity, assume that D4(D4')-branes are separated from each other.
This corresponds in gauge theory language to turning on the vev of the adjoint hypermultiplet 
scalars, and therefore the gauge group $U(Q_1)\times U(Q_2)$ is broken to $U(1)^{Q_1}\times 
U(1)^{Q_2}$.
As we have seen in the $Q_1=Q_2=1$ case, BPS monopoles will be described by the gauge fields 
and the bi-fundamental hypermultiplet scalars.
The relevant part of the effective theory action is therefore as follows,
\begin{eqnarray}
       S &=& \int d^3x [\sum_{n=1}^{Q_1}(-\frac1{4g^2}F_{nij}^{(1)}F_n^{(1)ij}-\frac{g^2}2 
              (D_n^{(1)})^2 -g^2|F_n^{(1)}|^2) \nonumber \\
         & & +\sum_{m=1}^{Q_2}(-\frac1{4g^2}F_{mij}^{(2)}F_m^{(2)ij}-\frac{g^2}2
              (D_m^{(2)})^2 -g^2|F_m^{(2)}|^2) \\
         & & -\sum_{n=1}^{Q_1}\sum_{m=1}^{Q_2}|D_iq_{nm}|^2], \hspace{1cm} (i,j=0,1,2)
               \nonumber
\end{eqnarray}
where,
$$ D_iq_{nm} = \partial_i q_{nm}+i(A_{ni}^{(1)}-A_{mi}^{(2)})q_{nm} $$
\begin{eqnarray}
& & D_n^{(1)} = -\sum_{m=1}^{Q_2}(|q_{nm}|^2-|\tilde{q}_{mn}|^2-\zeta) \nonumber \\
& & D_m^{(2)} = +\sum_{n=1}^{Q_1}(|q_{nm}|^2-|\tilde{q}_{mn}|^2-\zeta) \\
& & F_n^{(1)} = -\sqrt2 \sum_{m=1}^{Q_2} q_{nm}\tilde{q}_{mn} \nonumber \\
& & F_m^{(2)} = +\sqrt2 \sum_{n=1}^{Q_1} \tilde{q}_{mn}q_{nm} .\nonumber
\end{eqnarray}
$A_{ni}^{(1)}$ and $A_{mi}^{(2)}$ are the gauge fields in the Cartan subalgebra of 
$U(Q_1)$ and $U(Q_2)$.
The FI-parameter $\zeta$ is introduced for the same reason as before.

First we consider the vacuum.
Suppose that $\zeta$ is positive.
The vanishinig of the potential leads to the vacuum configuration.
\begin{equation}
|q_{nm}| = \zeta ,\hspace{5mm} \tilde{q}_{mn} = 0
\end{equation}
The monopole solutions must satisfy this conditions at spatial infinity.

Then we investigate the static BPS monopole solutions which preserve half of 
the supersymmetries.
The staic energy of the system is
\begin{eqnarray}
E &=& \int d^2x [\sum_{n=1}^{Q_1}(\frac1{4g^2}F_{n\alpha\beta}^{(1)}F_{n\alpha\beta}^{(1)}
                 +\frac{g^2}2 (D_n^{(1)})^2) 
             +\sum_{m=1}^{Q_2}(\frac1{4g^2}F_{m\alpha\beta}^{(2)}F_{m\alpha\beta}^{(2)}
                 +\frac{g^2}2(D_m^{(2)})^2 ) \nonumber \\
         & & +\sum_{n=1}^{Q_1}\sum_{m=1}^{Q_2}|D_\alpha q_{nm}|^2 ],\hspace{1cm}
               ( \alpha,\beta=1,2 )
\end{eqnarray}
where we set $\tilde{q}_{mn}=0$ everywhere.
This can be rewritten as follows,
\begin{eqnarray}
E &=& \int d^2x [ \sum_{n=1}^{Q_1}\frac1{2g^2}
        (\frac12\epsilon_{\alpha\beta} F_{n\alpha\beta}^{(1)}\pm g^2D_n^{(1)})^2
                 +\sum_{m=1}^{Q_2}\frac1{2g^2}
        (\frac12\epsilon_{\alpha\beta} F_{m\alpha\beta}^{(2)}\pm g^2D_m^{(2)})^2 \nonumber \\
  & & +\sum_{n=1}^{Q_1}\sum_{m=1}^{Q_2}
        \frac12|D_\alpha q_{nm}\pm i\epsilon_{\alpha\beta}D_\beta q_{nm}|^2 ] \\
  & & \mp\frac12\zeta \sum_{n=1}^{Q_1}\sum_{m=1}^{Q_2}
           \int d^2x \epsilon_{\alpha\beta}(F_{n\alpha\beta}^{(1)}-F_{m\alpha\beta}^{(2)})
        \nonumber
\end{eqnarray}
where we used that $D_\alpha q_{nm}\to 0$ at spatial infinity.
Thus the energy is bounded from below.
The BPS equations are then,
\begin{eqnarray}
& & \frac12\epsilon_{\alpha\beta} F_{n\alpha\beta}^{(1)}\pm g^2D_n^{(1)}=0 \nonumber \\
& & \frac12\epsilon_{\alpha\beta} F_{m\alpha\beta}^{(2)}\pm g^2D_m^{(2)}=0 \label{BPSeq} \\
& & D_\alpha q_{nm}\pm i\epsilon_{\alpha\beta}D_\beta q_{nm}=0. \nonumber
\end{eqnarray}
These equations are the generalization of the well-known Nielsen-Olesen vortex equations 
\cite{vortex}.
One can verify that the solutions to (\ref{BPSeq}) preserve half of the supersymmetries,
if we choose the same signs for all $n$ and $m$.

\vspace{7mm}

3.1 \ The simplest case: $Q_1=2,Q_2=1$

\bigskip

We will consider the simplest case in which there are two D4-branes and a D4'-brane, thus 
there are two intersections.
In the following, we will take the lower sign in (\ref{BPSeq}), which corresponds to 
investigating the positive-charge monopole solutions.
The BPS equations are the following.
\begin{eqnarray}
& & F_{1,12}^{(1)} = g^2(-|q_{11}|^2+\zeta) \\
& & F_{2,12}^{(1)} = g^2(-|q_{21}|^2+\zeta) \\
& & F_{1,12}^{(2)} = g^2(|q_{11}|^2+|q_{21}|^2-2\zeta) \\
& & \hspace*{-2cm} (\partial_1 q_{11}+i(A_{1,1}^{(1)}-A_{1,1}^{(2)})q_{11})
-i(\partial_2 q_{11}+i(A_{1,2}^{(1)}-A_{1,2}^{(2)})q_{11}) = 0 \\
& & \hspace*{-2cm} (\partial_1 q_{21}+i(A_{2,1}^{(1)}-A_{1,1}^{(2)})q_{21})
-i(\partial_2 q_{21}+i(A_{2,2}^{(1)}-A_{1,2}^{(2)})q_{21}) = 0
\end{eqnarray}
We redefine the fields as follows.
\begin{eqnarray}
A_\alpha = A_{1,\alpha}^{(1)}-A_{1,\alpha}^{(2)},& & \hspace{3mm} 
\tilde{A}_\alpha = A_{2,\alpha}^{(1)}-A_{1,\alpha}^{(2)} \nonumber \\
F = \partial_1 A_2-\partial_2 A_1,& & \hspace{3mm} 
\tilde{F} = \partial_1 \tilde{A}_2-\partial_2 \tilde{A}_1 \\
q = q_{11},\hspace{1.8cm}& & \hspace{3mm} \tilde{q} = q_{21} \nonumber
\end{eqnarray}
After suitable rescaling, we obtain 
\begin{eqnarray}
& & \hspace{1cm} F = -2|q|^2-|\tilde{q}|^2+3 \label{eq1} \\
& & \hspace{1cm} \tilde{F} = -|q|^2-2|\tilde{q}|^2+3 \\
& & (\partial_1q+iA_1q)-i(\partial_2q+iA_2q) = 0 \\
& & (\partial_1\tilde{q}+i\tilde{A}_1\tilde{q})-i(\partial_2\tilde{q}+i\tilde{A}_2\tilde{q})
= 0 \label{eq2}
\end{eqnarray}
The original gauge fields are determined by the use of the condition
\begin{equation}
F_{1,12}^{(1)}+F_{2,12}^{(1)}+F_{1,12}^{(2)}=0.
\end{equation}

First we discuss the topological properties of the solutions.
The boundary conditions at infinity ($|q|\to 1,D_\alpha q\to 0$) lead
\begin{equation}
q\to e^{-i\varphi}, \hspace{3mm} A_\alpha\to \partial_\alpha \varphi.
\end{equation}
Then the magnetic charge is
\begin{equation}
\frac1{2\pi}\int d^2x F = \frac1{2\pi}(\varphi(\theta=2\pi)-\varphi(\theta=0)). 
\label{charge}
\end{equation}
For the single-valuedness of $q$, the RHS of (\ref{charge}) is an integer $n$.
Similar result holds for $\tilde{q},\tilde{A}_\alpha$ and its magnetic charge is another 
integer $\tilde{n}$.
The energy of this solution is
\begin{equation}
E=\frac12\zeta(n+\tilde{n})
\end{equation}
This shows that the solution is stable due to its topological property.

Next we consider the radially symmetric solutions with magnetic charge $n$ and $\tilde{n}$.
Suppose the following ansatz.
\begin{eqnarray}
q = f(r)e^{-in\theta},\hspace{3mm} 
A_\alpha = -\epsilon_{\alpha\beta}x_\beta \frac{\tilde{n}}{r^2}a(r) \\
\tilde{q} = \tilde{f}(r)e^{-i\tilde{n}\theta},\hspace{3mm}
\tilde{A}_\alpha = -\epsilon_{\alpha\beta}x_\beta\frac{\tilde{n}}{r^2}\tilde{a}(r)
\end{eqnarray}
($r,\theta$) is the polar coordinates.
This reduces (\ref{eq1})$\sim$(\ref{eq2}) to the ordinary differential equations.
\begin{eqnarray}
r\frac{df}{dr} = n(1-a)f,\hspace{3mm} \label{global}
\frac nr \frac{da}{dr} = -2f^2-\tilde{f}^2+3 \\
r\frac{d\tilde{f}}{dr} = \tilde{n}(1-\tilde{a})\tilde{f},\hspace{3mm}
\frac{\tilde{n}}r \frac{d\tilde{a}}{dr} = -f^2-2\tilde{f}^2+3
\end{eqnarray}
The boundary conditions are then
\begin{equation}
f,\tilde{f}\to 1,\hspace{3mm} a,\tilde{a}\to 1\hspace{3mm} (r\to \infty). \label{boundary}
\end{equation}

Now we argue the global property of the solutions.
It is natural to expect that for the positive-charge monopole solutions $F,\tilde{F}\geq 0$, 
where $F=\frac nr\frac{da}{dr}$ etc.
This means that $a,\tilde{a}$ are monotonically increasing functions.
From (\ref{boundary}), we can see that $a,\tilde{a}\leq 1$.
Integrating (\ref{global}) and use the boundary value of $f$, we obtain
\begin{equation}
\log f(r) = -\int_r^\infty d\rho \frac1\rho n(1-a(\rho))
\end{equation}
The RHS is always negative, indicating that $0\leq f \leq 1$.
Especially, since $a(0)< 1$ the RHS must diverge at $r=0$.
This means $f(0)=0$.
Similar argument leads to $\tilde{a}(0)<1,\tilde{f}(0)=0$.

Such a solution can exist around $r=0$.
It is useful to define $h(r)=n(1-a(r))$ etc. The solution is 
\begin{eqnarray}
& & f(r) = cr^n +o(r^{n+1}) \\
& & \tilde{f}(r) = \tilde{c}r^{\tilde{n}} +o(r^{\tilde{n}+1}) \\
& & h(r) = n-\frac32 r^2 +o(r^{2(m+1)}) \\
& & \tilde{h}(r) = \tilde{n}-\frac32 r^2+o(r^{2(m+1)}) ,
\end{eqnarray}
where $m=\mbox{max}(n,\tilde{n})$.
$c,\tilde{c}$ are non-zero constants.

Assuming that such solutions exist globally, one can calculate the dimension of the moduli 
space ${\cal M}_{n,\tilde{n}}$ of the monopole solutions with charges ($n,\tilde{n}$).
From the ordinary index calculation, we obtain
\begin{equation}
\mbox{dim}_{\bf R}{\cal M}_{n,\tilde{n}} = 2(n+\tilde{n})
\end{equation}
This can be interpreted as follows.
Remember that the system considered here is the brane configuration with two intersections.
Therefore the solutions with charges ($n,\tilde{n}$) correspond to the configuration 
in which $n$ D0-branes are on one intersection, and $\tilde{n}$ D0-branes are on another 
intersection.
The moduli will correspond to the positions of D0-branes on each intersection.

We expect that the above arguments can be extended to the $Q_1\geq 3,Q_2=1$ case.

\hspace{7mm}

3.2 \ $Q_1=Q_2$ case

\bigskip

Now we consider more general case: $Q_1=Q_2=Q$.
The properties of the solutions will be similar to the previous case.
From the single-valuedness of $q_{nm}$, the magnetic charges are quantized,
\begin{equation}
\frac1{2\pi}\int d^2x F_{nm} = l_{nm} \in {\bf Z}
\end{equation}
where
\begin{equation}
F_{nm} = F_{n,12}^{(1)} - F_{m,12}^{(2)}.
\end{equation}
One can easily see that all $l_{nm}$'s are not independent.
They obey
\begin{equation}
l_{nm}+l_{n+1,m+1}=l_{n+1,m}+l_{n,m+1}. \label{constraint}
\end{equation}
Then the simplest charge distributions which satisfy (\ref{constraint}) are, for $Q=2$,
\begin{equation}
\left( \begin{array}{cc} l_{11} & l_{12} \\ l_{21} & l_{22} \end{array} \right)
= \left( \begin{array}{cc} 1 & 0 \\ 1 & 0 \end{array} \right),\hspace{3mm}
  \left( \begin{array}{cc} 1 & 1 \\ 0 & 0 \end{array} \right) \hspace{3mm}
  \mbox{etc.} \label{solutions}
\end{equation}
The D0-brane interpretation of the solutions (\ref{solutions}) are as follows.
Consider again the original non-Abelian theory.
The number of the D0-branes $N$ should be given as
\begin{equation}
N = \frac1{2\pi}\int d^2x (trF_{12}^{(1)} - trF_{12}^{(2)}),
\end{equation}
($trF_{12}^{(1)}+trF_{12}^{(2)}=0$ for the BPS solutions).
Thus in the Abelian theory discussed above the number of the D0-branes is given as
\begin{eqnarray}
N &=& \frac1{2\pi}\int d^2x (\sum_{n=1}^Q F_{n,12}^{(1)}-\sum_{m=1}^Q F_{m,12}^{(2)}) 
       \nonumber\\
  &=& \frac1{2\pi}\int d^2x \sum_{n=1}^Q F_{nn} \nonumber \\
  &=& \sum_{n=1}^Q l_{nn}.
\end{eqnarray}
The distributions (\ref{solutions}) then corresponds to $N=1$.
General solutions will be obtained as the superpositions of such simple solutions, and 
there will exist the solution for all $N$.
Therefore the BPS monopole solutions can be interpreted as the D0-branes on the 
intersections.

\vspace{7mm}

{\large {\bf 4. Discussions}}

\bigskip

We have constructed the effective thoery on the D4-D4' intersections in which all 
D4(D4')-branes are separated from each other, and studied the properties of the BPS monopole 
solutions in the theory.
The solutions allow us to interpret them as the D0-branes on the intersections.
In $Q_1=2,Q_2=1$ case, there is evidence for the existence of the monopole solutions, 
and its moduli will correspond to the positions of D0-branes on the intersections.
In more general cases, it is argued that there may exist the monopoles corresponding to any 
number of the D0-branes.

In contrast to the case in the previous paper \cite{suyama}, the derivation of the entropy 
by counting the ground states of the BPS monopoles is less straightforward in the present 
context.
As we have seen in the $Q_1=Q_2=2$ case, the ``unit" solutions will be (\ref{solutions}).
It is hard to interpret them as a D0-brane on a particular intersection.
This means that the number of states which contribute to the entropy might be less than what 
is expected.
We think that these missing states are recovered, because the moduli space of the ``unit" 
solution (\ref{solutions}) will be larger than the one obtained in the $Q_1=Q_2=1$ case, 
and these solutions will have larger number of ground states than one might think.

Thus we can think that the description discussed in this paper is an intermediate step 
toward the construction of quantum mechanics of the black hole.
When we consider the non-Abelian theory, the only quantum number that labels BPS states 
will be the number of the D0-branes.
Then all solutions are included in a moduli space and cannot be counted individually.
It is possible, in principle, to construct quantum mechanics of the black hole as 
$\sigma$-model whose target space is the moduli space of the BPS monopole solutions.
The microstates which contribute to the entropy will be obtained as the ground states, 
which corresponds to the cohomology elements of the moduli space.
Moreover the dynamics of the $\sigma$-model will describe the physics of the near-extremal 
black hole.

\hspace{7mm}

{\bf Acknowledgment}

\bigskip

I would like to thank H.Itoyama, K.Murakami, T.Yokono for valuable discussions.


\newpage

\begin{thebibliography}{99}

\bibitem{SV}A.Strominger, C.Vafa, {\it Microscopic Origin of Bekenstein-Hawking Entropy}, 
Phys. Lett. {\bf B379} (1996) 99, hep-th/9601029.
\bibitem{CM}C.Callan, J.Maldacena, {\it D-brane Approach to Black Hole Quantum Mechanics}, 
Nucl. Phys. {\bf B472} (1996) 591, hep-th/9601029.
\bibitem{BH}J.Bekenstein, {\it Black Holes and Entropy}, Phys. Rev. {\bf D7} (1973) 2333; \\
S.Hawking, {\it Particle Creation by Black Holes}, Comm. Math. Phys. {\bf 43} (1975) 199. 
\bibitem{solution}M.Cvetic, D.Youm, {\it Dyonic BPS Saturated Black Holes of Heterotic 
String on a Six-Torus}, Phys. Rev. {\bf D53} (1996) 584, hep-th/9507090; \\
A.Tseytlin, {\it Extreme dyonic black holes in string theory}, Mod. Phys. Lett. {\bf A11} 
(1996) 689, hep-th/9601177; \\
I.Klebanov, A.Tseytlin, {\it Intersecting M-branes as Four-Dimensional 
Black Holes}, Nucl. Phys. {\bf B475} (1996) 179, hep-th/9604166.
\bibitem{4dim1}J.Maldacena, A.Strominger, {\it Statistical Entropy of Four-Dimensional 
Extremal Black Holes}, Phys. Rev. Lett. {\bf 77} (1996) 428, hep-th/9603060.
\bibitem{4dim2}C.Johnson, R.Khuri, R.Myers, {\it Entropy of 4D Extremal Black Holes}, 
Phys. Lett. {\bf B378} (1996) 78, hep-th/9603061. \\
G.Horowitz, D.Lowe, J.Maldacena, {\it Statistical Entropy of Nonextremal Four- Dimensional 
Black Holes and U-Duality}, Phys. Rev. Lett. {\bf 77} (1996) 430, hep-th/9603195.
\bibitem{BinB}M.Douglas, {\it Branes within Branes}, hep-th/9512077.
\bibitem{reduction}J.Maharana, J.Schwarz, {\it Noncompact Symmetries in String Theory}, 
Nucl.Phys. {\bf B390} (1993) 3, hep-th/9207016; \\
A.Sen, {\it Electric Magnetic Duality in String Theory}, Nucl. Phys. {\bf B404} (1993) 109, 
hep-th/9207053.
\bibitem{suyama}T.Suyama, {\it Monopoles and Black Hole Entropy}, hep-th/9909091.
\bibitem{intersection}A.Sen, {\it U-duality and Intersecting D-branes}, 
Phys. Rev. D53 (1996) 2874; \\
A.Hanany, I.Klebanov, {\it On Tensionless Strings in 3+1 Dimensions}, 
 Nucl. Phys. B482 (1996) 105.
\bibitem{vortex}H.Nielsen, P.Olesen, {\it Vortex Line Models for Dual Strings}, 
Nucl. Phys. {\bf B61} (1973) 45; \\
H.de Vega, F.Schaposnik, {\it Classical Vortex Solution of the Abelian Higgs Model}, 
Phys. Rev. {\bf D14} (1976) 1100; \\
E.Weinberg, {\it Multivortex Solutions of the Ginzburg-Landau Equations}, 
Phys. Rev. {\bf D19} (1979) 3008; \\
C.Taubes, {\it Arbitrary N-Vortex Solutions to the First Order Ginzburg-Landau Equations}, 
Comm. Math. Phys. {\bf 72} (1980) 277.


\end{thebibliography}
\end{document}